%% file: main.tex
\documentclass[acmsmall]{acmart}

\usepackage[utf8]{inputenc}

\PassOptionsToPackage{table,xcdraw}{xcolor} 

\usepackage{makecell}
\usepackage{textgreek}
\usepackage{multirow}
\usepackage{subcaption}
\usepackage{xcolor,colortbl}
\definecolor{gray}{rgb}{0.1,0.1,0.1}
\usepackage[T1]{fontenc}
\usepackage{graphicx}
\usepackage{tabularx}
\usepackage{longtable}
\usepackage{enumitem}
\usepackage{booktabs}
\usepackage{array}
\usepackage{setspace}

\AtBeginDocument{%
  \providecommand\BibTeX{{%
    \normalfont B\kern-0.5em{\scshape i\kern-0.25em b}\kern-0.8em\TeX}}}

\acmConference[Woodstock '18]{Woodstock '18: ACM Symposium on Neural
  Gaze Detection}{June 03--05, 2018}{Woodstock, NY}
\acmBooktitle{Woodstock '18: ACM Symposium on Neural Gaze Detection,
  June 03--05, 2018, Woodstock, NY}
\acmPrice{15.00}
\acmISBN{978-1-4503-XXXX-X/18/06}

\acmSubmissionID{123-A56-BU3}

\input{sections/header}

\begin{document}
\title[{Social Media Should Feel Like Minecraft, Not Instagram:\\Youth Visions for Meaningful Social Connections through Fictional Inquiry}]{Social Media Should Feel Like Minecraft, Not Instagram: Youth Visions for Meaningful Social Connections through Fictional Inquiry}

\author{JaeWon Kim}
\orcid{0000-0003-4302-3221}
\affiliation{%
  \institution{University of Washington}
  \city{Seattle}
  \state{WA}
  \country{USA}}
\email{jaewonk@uw.edu}

\author{Hyunsung Cho}
\orcid{0000-0002-4521-2766}
\affiliation{%
  \institution{Human-Computer Interaction Institute, Carnegie Mellon University}
  \city{Pittsburgh}
  \country{USA}}
\email{hyunsung@cs.cmu.edu}

\author{Fannie Liu}
\orcid{0000-0002-5656-3406}
\affiliation{%
  \institution{JPMorgan Chase \& Co.}
  \city{New York}
  \state{NY}
  \country{USA}
}
\email{fannie.liu@jpmchase.com}

\author{Alexis Hiniker}
\orcid{0000-0003-1607-0778}
\affiliation{%
  \institution{The Information School, University of Washington}
  \city{Seattle}
  \country{USA}}
\email{alexisr@uw.edu}

\renewcommand{\shortauthors}{JaeWon Kim, et al.}

\begin{abstract}
We conducted co-design workshops with 23 participants (ages 15--24) to explore how youth envision ideal remote social connection. Using Fictional Inquiry (FI) within a Harry Potter-inspired narrative, we found that youth perceive a disconnect between platforms labeled ``social media'' (like Instagram) and those where they actually experience meaningful connections (like Minecraft or Discord). Participants envisioned an immersive platform prioritizing meaningful social connection through presence and immersion, natural self-expression, intuitive spatial navigation leveraging physical-world norms, and playful, low-stakes opportunities for friendship development. We synthesize these visions into six themes articulating relational needs that current platforms systematically marginalize. The FI method proved effective in generating innovative ideas while empowering youth by fostering hope and agency over social media's future. Our findings challenge ``doom'' narratives by reframing social media's harms as outcomes of specific design choices, demonstrating how design research can reopen space for imagining more supportive forms of mediated connection.
\end{abstract}

\ccsdesc[500]{Human-centered computing~Collaborative and social computing}

\keywords{social media; fictional inquiry; youth; design}

\maketitle

\input{sections/1_introduction}
\input{sections/2_related-work}
\input{sections/3_method}
\input{sections/4_results}
\input{sections/5_discussion}
\input{sections/6_conclusion}


\bibliographystyle{ACM-Reference-Format}
\bibliography{references, references2}

\input{sections/7_appendix}

\end{document}

%% file: sections/header.tex
\input{formatting/published-packages} 
\input{formatting/published-commands} 


%% file: formatting/published-packages.tex
\usepackage{xspace}     
\usepackage{xpunctuate} 

%% file: formatting/published-commands.tex
\newcommand{\inlinequote}[2][]{
  \emph{``#2''}%
  \ifx&#1&%
  \allowbreak%
  \else%
  \ (P#1)\allowbreak%
  \fi%
}

\newcommand{\blockquote}[2]{           
    \begin{quote}\emph{``#1''} (P#2)\end{quote}\allowbreak}



%% file: sections/1_introduction.tex
\section{Introduction}
Meaningful social connections are critical for young people's development and well-being~\cite{Hawkley-2010-LonelinessMattersMechanisms-v}. Early social computing researchers and users celebrated these technologies for their potential to build relationships that transcend physical boundaries~\cite{ellison2007benefits}. Over time, however, that promise has narrowed. Between 2023 and 2025, time spent viewing friends' content dropped from 22\% to 17\% on Facebook and from 11\% to 7\% on Instagram~\cite{Bradley-2025-InstagramFacebookProof-l}. Current designs increasingly prioritize passive consumption over active interaction with close friends~\cite{Landesman-2024-IInstagram-j}, and are associated with fatigue, mistrust, and perceived harm. Platforms once imagined as infrastructures for personal connection now make even small acts of self-disclosure---the basic currency of relationship development---feel high-stakes~\cite{Kim2025Trust-EnabledRegulationz} and anxiety-inducing~\cite{jaewon-privacy}.

For youth who have grown up with these technologies as central to their social lives, this shift carries consequences beyond user experience. While young people report frustration, anxiety, and toxicity, they also describe moments of genuine joy and meaningful connection---and participation remains a practical necessity for staying connected with peers. Yet dominant policy, educational, and design responses frame youth engagement in terms of prevention and protection, emphasizing restriction over support for experimentation that might foster agency and resilience~\cite{Kim2025DesignChallengese, Kim2024EnvisioningPreventingn}. Youth consistently report that measures such as outright bans leave them feeling unheard~\cite{Kim-2024-AustraliaBarred-y, TheLearningNetwork-2025-WhatTeensMedia-z, Weinstein2022-rh}. Their relationships with social media are ambivalent rather than uniformly negative, but this complexity is rarely reflected in prevailing approaches to governance or design.

This disconnect is reinforced by the broader convergence of social media platforms themselves~\cite{Sundaram-Other-SocialMediaSame-g, Pardes-2020-SocialMediaSame-p, Pitt-2021-SocialMediaSame-s, Nielsen-2023-ComparingPlatformizationAnalysis-m, VanDyke-2020-GrowingSimilaritiesPlatforms-m}. Despite surface-level differences, mainstream platforms increasingly share a common paradigm organized around feeds, mobile interfaces, and vertical profiles, crowding out alternative models of mediated connection. As a result, much design research now operates in a reactive mode. The guiding question has shifted from what meaningful online connection might look like to how people can be protected from existing platforms. This framing treats the dominant paradigm as fixed, focusing on mitigation rather than reimagination. Although generative work exists within HCI and related fields, it is often overshadowed by protectionist narratives that position youth as vulnerable users in need of safeguards, prioritizing safety over empowerment and incremental repair over structural alternatives~\cite{Kim2025DesignChallengese, Kim2024EnvisioningPreventingn}.

These constraints shape not only policy and design discourse but also participatory research itself. When young people are asked to redesign social media, the prompt arrives burdened with existing controversies around addiction, mental health, and toxicity. Participants often respond defensively, reproducing adult-driven narratives through social desirability bias even when their lived experiences are more nuanced~\cite{Davis2019EverythingsRelationshipsg}. Without language or precedents beyond existing platforms, participants tend to remix familiar features---a cleaner feed, better moderation, fewer metrics---rather than imagining fundamentally different forms of social interaction. The term ``social media'' itself anchors imagination to platforms like Instagram or TikTok, constraining what can be envisioned before design even begins. As a result, participatory design risks generating refinements when its prompts and methods keep participants inside the same conceptual frame. To move beyond these constraints, we ask:

\begin{itemize}
    \item \textit{\textbf{RQ:} What forms of social interaction do youth envision when freed from current platform constraints?}
\end{itemize}
Rather than beginning from existing systems or problems, we focus on the relationships, interactions, and forms of presence young people desire, and on the design considerations that emerge from these visions.

To address this question, we conducted co-design interviews using the Fictional Inquiry (FI) method~\cite{IversenDindler-2007-FictionalInquiry--designSpace-m}, a speculative approach that invites participants to imagine within a fictional scenario removed from real-world constraints. Twenty-three participants aged 15–24 imagined how students at Hogwarts---the fictional school from \textit{Harry Potter}\cite{harrypotterfanclub}---might connect with friends beyond their campus using any magical powers they could envision. This framing served three purposes. First, it functioned as a defamiliarization strategy, breaking the conceptual lock of existing platforms by removing both technical limitations and loaded terminology, starting from desire rather than constraint. Second, it centered meaningful social connection as the explicit design goal, reflecting youth developmental needs and interests documented in prior research~\cite{Davis2012-bq}. Third, it created a space of play in which youth are natural experts, enabling them to express values, social logics, and relational priorities difficult to articulate within dominant narratives about social media. Participants were not asked to redesign social media; they were asked how wizards might stay close across distance.

Our findings reveal six themes characterizing how youth envision meaningful remote social interaction: intuitive social navigation modeled on real-world movement and presence; shared experiences through collaborative activities and casual co-presence; communal environments that prioritize close relationships and low-pressure interaction; flexible self-presentation through avatars and contextual identity management; intentional engagement that replaces passive scrolling with purposeful navigation; and playful, low-stakes social mechanics. Within these themes, participants described concrete interactions---sitting together in a virtual study room with Lo-Fi music, leaving a small gift at a friend's door without initiating a full conversation, or using secret knocks to grant trusted friends access to private spaces. Although participants developed their visions independently, their accounts of how these imagined systems would feel converged strongly, emphasizing connection, engagement, fulfillment, and agency. These patterns suggest shared relational needs not well supported by current platforms. We also observed that the FI process itself shifted participants from reactive problem-fixing toward aspirational imagining, with many reporting increased hope and excitement about what mediated connection could become.

This paper makes three contributions. First, we offer a methodological account of FI as a research-through-design approach for reimagining meaningful remote social connection. By suspending constraints and removing the term ``social media'' itself, the fictional framing enabled youth---a population often positioned as subjects to protect rather than agents of imagination---to reason from felt experience rather than inherited platform templates. Second, we present an empirical account of the interaction concepts, affordances, and social mechanisms young people proposed when freed from feed-based paradigms. Third, we synthesize these concepts into six themes that articulate underlying relational needs and design directions, highlighting what modern social media systematically marginalizes. Together, these contributions challenge ``doom'' narratives around social media, reframing them as outcomes of specific design choices and demonstrating how design research can reopen space for imagining more supportive forms of mediated connection.

%% file: sections/2_related-work.tex
\section{Related Work}

\subsection{Beyond Moral Panic: From Panic Narratives to Design Opportunity}
Widespread societal concerns exist about social media's impact on youth well-being~\cite{OtherOther-2023-SocialMediaAdvisory-u}. Rather than addressing these constructively, public discourse has framed social media as a societal evil~\cite{Wike-2022-2ViewsSociety-y}, particularly in the U.S., where it is blamed for issues ranging from democratic instability to what some call the \inlinequote{Cause of the Mental Illness Epidemic}~\cite{Haidt-2023-SocialMediaEvidence-c}. This narrative has prompted restrictive policy responses worldwide, such as the \inlinequote{Stop Addictive Feeds Exploitation (SAFE) For Kids Act.}

However, treating social media as inherently harmful, or youth as best served by removal, collapses a more complex reality. Despite public concern about social media's role in youth mental health, causal evidence remains inconclusive, and youth report a mix of positive and negative experiences~\cite{Vogels-2023-TeensSocialSurveys-n}. Fear-based policies and alarmist narratives can themselves cause harm, amplifying anxiety, mistrust, and helplessness among youth~\cite{kim2024privacysocialnormsystematically, lee2024social}. Recognizing nuance does not deny that serious harms occur, including high-risk vulnerabilities (e.g., sexual exploitation, grooming, and synthetic media threats), nor ignore structural forces like platform power asymmetries and algorithmic manipulation. Instead, it motivates a complementary design question: what kinds of online social environments could better support youth relationships while accounting for both everyday experiences and real risks?

\subsection{Meaningful Social Connection Online, and Why Mainstream Platforms Miss It}
\paragraph{What counts as meaningful connection}
Litt et al.~\cite{Litt-2020-WhatMeaningfulSurvey-h} define meaningful social interactions as \inlinequote{interactions that people believe enhance their lives, the lives of their interaction partners, or their relationships, with emotional, informational, or tangible impact.} Though \inlinequote{deeply subjective,} these interactions are distinguished by their impact \inlinequote{beyond the immediate interaction itself.} They identify three key impact types: \inlinequote{emotional impact} (feelings, mood changes, or relationship strengthening), \inlinequote{informational impact} (knowledge, advice, or understanding exchange), and \inlinequote{tangible impact} (concrete benefits like goods or services). Strong ties, community ties, and shared activities facilitate meaningfulness, while attributes like advance planning, memorialization, and synchronicity often support impact.

Building on this work, we treat meaningful social interaction as any interaction participants would actively seek in an ideal scenario, reflecting their subjective sense of meaningfulness.

\paragraph{Connection-seeking on platforms optimized for other gratifications}
Though social media platforms are primary connection tools for young people~\cite{BoydBoyd-2014-ComplicatedSocialTeens-j}, many users report feeling lonely or disconnected despite widespread use~\cite{twenge2019less, primack2017social}, highlighting a gap between social media's promise and lived experience. Mainstream platforms like Instagram~\cite{instagram}, YouTube~\cite{youtube}, and Discord~\cite{discord} serve diverse user motivations including entertainment, information seeking, and self-promotion~\cite{whiting2013people, Menon-2022-UsesGratificationsInstagram-x}. From a uses and gratifications perspective~\cite{GurevitchKatz-1973-UsesGratificationsResearch-y}, users pursue varied goals from building social capital~\cite{phua_uses_2017} to media consumption~\cite{Menon-2022-UsesGratificationsInstagram-x}. Users who primarily seek meaningful connection may thus encounter experiences that prioritize other forms of engagement, leaving social needs unmet.

HCI and CSCW work has explored designs supporting more meaningful interaction, including efforts encouraging effortful communication~\cite{LiuFannie2021SOUt, LiuZhang-2022-AuggieEncouragingExperiences-n} and reflective or value-oriented social interaction~\cite{Stepanova2022-vh}. More broadly, recent work articulates design spaces and system-level alternatives beyond dominant templates (e.g., form factors, interaction conventions, and social architectures)~\cite{zhang2024form}. Yet if connection is so central, why do mainstream platforms consistently fail to prioritize it? Part of the answer lies in structural forces constraining both platform incentives and design imagination.

\subsection{Recognizing and Overcoming Social Media Design Fixation}
Design fixation~\cite{jansson1991design} describes the tendency to remain constrained by conventional thinking or existing solutions, even when divergent exploration is needed. Social media design faces similar fixation~\cite{Sundaram-Other-SocialMediaSame-g, Pardes-2020-SocialMediaSame-p, Pitt-2021-SocialMediaSame-s, Nielsen-2023-ComparingPlatformizationAnalysis-m, VanDyke-2020-GrowingSimilaritiesPlatforms-m}. Ad-based revenue models incentivize increased time on apps~\cite{Other-Other-PlatformCapitalism-f}, shaping platforms toward engagement metrics and features that prolong attention. Platforms deploy compulsive strategies like friend recommendations~\cite{HinikerKim-2024-SharingDesign-x} and engagement metrics (e.g., Snap scores~\cite{rozgonjuk2021comparing}), which erode trust and foster distrustful user communities~\cite{Kim2025Trust-EnabledRegulationz}. In parallel, design efforts are often dominated by reactive harm mitigation rather than proactive exploration of new approaches~\cite{Kim2024EnvisioningPreventingn}. Together, these forces help explain why users experience mainstream platforms as converging toward similar, feed-driven patterns, even when those patterns do not align with meaningful connection goals.

These constraints motivate methods that deliberately create distance from current systems, enabling participants to articulate what they would want social media to optimize for without starting from existing platform assumptions.

\subsection{Alternative Paradigms and Methods for Imagining Beyond Platform Templates}
\paragraph{Sociality in games and multi-space online communities}
Video games are increasingly recognized as valuable environments for meaningful social connection, particularly among youth. Large-scale surveys show games serve as major venues for friendship development and maintenance~\cite{Lenhart-2015-Chapter3Boys-v}, with teens playing with known friends and making new ones through online gameplay~\cite{Lenhart-2015-Chapter1Friendships-i}. CSCW research shows that gaming communities often extend across multiple spaces, requiring coordination and group management across platforms and contexts~\cite{toombs2022we}. These findings suggest gaming-influenced sociality can illuminate how people sustain relationships through shared activities, low-stakes co-presence, and community infrastructure, while raising questions about what translates and who may be excluded by more spatial or higher-cognitive-load interaction styles.

\paragraph{Spatial social technologies and place-based interaction}
Beyond archetypal social media, HCI and CSCW have examined spatial social technologies emphasizing immersive and embodied interaction. Social VR and virtual worlds support co-presence, shared experiences, and identity play through avatars and navigable spaces~\cite{Zamanifard-2023-SurpriseBirthdayDistance-n, Freeman-2024-MyAudiences-u, Freeman-2020-MyBodyReality-l, Zamanifard-2019-TogethernessCraveRelationships-s, deighan2023social, rzeszewski2024social}. Platforms like Gather.town~\cite{duarte2023experience,tu2022meetings} and VRChat provide place-like environments where spatial layout, navigation, and ambient co-presence shape interaction rhythms. A longer history of virtual worlds (e.g., Second Life) highlights that spatiality can enable rich social experience while introducing familiar challenges around governance, safety, and uneven participation. We draw on these literatures as precedents for treating online social spaces as places, where constructs from place theory, environmental psychology, and presence research (e.g., propinquity, wayfinding, place attachment, social presence, and co-presence) can help interpret how spatial organization relates to social connection.

\paragraph{Youth-centered design and fiction-based methods for escaping fixation}
Youth-focused HCI has long argued for engaging young people as stakeholders and co-designers rather than solely as subjects of protection. HCI has also developed fiction-based and speculative methods that create distance from present-day systems, making it easier to reason about values, tradeoffs, and alternative social arrangements. Related approaches include design fiction, speculative design, and pastiche scenarios, which use crafted narratives and artifacts to surface new possibilities and critique existing assumptions~\cite{blythe2006pastiche, dunneraby2013speculative}. Prior HCI work has drawn explicitly on fictional worlds, including Harry Potter, to design for social values and family coordination~\cite{brown2007locating}. 

Building on this methods lineage, we use FI to structure analogical transfer through a shared fictional narrative; unlike design fiction or pastiche scenarios that often center researcher-authored speculative artifacts, FI treats a well-known fictional world as a common design constraint and supports participant-led world-building of social architectures and norms. This approach is especially relevant for domains like social media, where incentive structures and design fixation can make it difficult to surface what youth might find meaningful outside current assumptions.

%% file: sections/3_method.tex
\section{Method}
We conducted co-design interviews with 23 participants aged 15--24 to examine how youth envision ideal social media platforms when freed from real-world constraints and fixation on archetypal platforms. All participants had prior experience with 3D games and were familiar with \textit{Harry Potter}~\cite{rowling2015harry, harrypotterfanclub}, a popular fantasy series featuring characters with magical powers. We used the FI method to scaffold participants' design ideation. All procedures were approved by the university's Institutional Review Board (IRB).

\subsection{FI as a Tool for Imaginative Design Exploration}
FI is an exploratory co-design method that enables participants to move beyond real-world constraints through immersion in semi-fictional contexts~\cite{IversenDindler-2007-FictionalInquiry--designSpace-m}. Through carefully crafted scenarios incorporating immersive elements such as physical artifacts or role-playing activities, participants take on generative roles, envisioning futuristic or fictional settings.

FI helps teams bypass existing sociocultural constraints by working within a shared, partially fictional narrative, which can make it easier to ask questions that might otherwise feel awkward and to examine specific issues in practice~\cite{IversenDindler-2007-FictionalInquiry--designSpace-m}. It also supports future-oriented thinking by reducing the perceived ``constraints of reality'' and shifting attention from fixing present-day problems toward imagining desirable alternatives~\cite{LeeHiniker-2017-Co-DesigningPreschoolersComicboarding-e}. More broadly, fictional settings function as a rigorous inquiry tool: they can surface new insights by contrasting fiction with reality, stress-test ideas through counterfactual reasoning, and reveal how imagined worlds shape real-world thinking and action~\cite{Schoeneborn2022FictionalInquiryg}.

Dindler and Iversen~\cite{IversenDindler-2007-FictionalInquiry--designSpace-m} outline three key implementation steps: defining the inquiry's purpose (e.g., staging design situations, exploring future ideas, or driving organizational change); developing a narrative that participants find approachable yet sufficiently distinct from current practices to encourage new thinking; and creating compelling plots that introduce tension or conflict to motivate design activity. Unlike traditional role-playing, FI encourages participants to retain their own identities and expertise while operating within fictional contexts, allowing them to explore new possibilities through personal motivations and capabilities while sidestepping societal constraints.

\subsubsection{Addressing Design Fixation Through FI}
We hypothesized that FI could support idea generation in social media design, a design space often perceived as stagnant or doomed. Perceptions of inevitability are reinforced by the convergence of mainstream platforms around similar structures and interaction conventions~\cite{Sundaram-Other-SocialMediaSame-g, Pardes-2020-SocialMediaSame-p, Pitt-2021-SocialMediaSame-s, Nielsen-2023-ComparingPlatformizationAnalysis-m, VanDyke-2020-GrowingSimilaritiesPlatforms-m}, as well as by economic incentives prioritizing attention capture and sustained time on platform~\cite{Other-Other-PlatformCapitalism-f}. Recent scholarship further suggests that design efforts have become dominated by reactive harm mitigation rather than proactive exploration of alternative paradigms~\cite{Kim2024EnvisioningPreventingn}. Together, these conditions constrain both imagination and inquiry, making it difficult to surface what people might find meaningful outside current platform assumptions.

By freeing participants from real-world constraints and encouraging them to draw on experiences from other media such as games, movies, VR, stories, and magical or fictional worlds, FI allowed participants to engage with concepts they find meaningful without concern for social desirability, ethical scrutiny, or technical feasibility. Given the strong design and cognitive fixation characterizing social media design, this form of analogical transfer may enable conceptual breakthroughs~\cite{jansson1991design,novick1988analogical}. Reframing social media design within a fictional setting also functions as a problem-reframing strategy, a core aspect of design thinking that can help designers move beyond entrenched assumptions~\cite{Buchanan-1992-WickedProblemsThinking-x}. Accordingly, we hypothesized that FI would be effective both in uncovering youths' underlying needs around remote social connection, free from dominant narratives or fear of judgment, and in enabling design breakthroughs through problem reframing and analogical transfer.

\subsection{Material: The Narrative Framework}
We selected the \textit{Harry Potter} world as our FI narrative framework. The series follows an orphaned boy who discovers he is a wizard and attends ``Hogwarts School of Witchcraft and Wizardry,'' forming deep friendships while uncovering his destiny and clashing with the antagonist.

Among numerous magical worlds, \textit{Harry Potter} proved particularly suitable for FI interviews. Its extensive popularity and cultural presence provide shared understanding, facilitating recruitment and engagement. By 2023, 98\% of US youth aged 18--27 identified as casual or avid fans, with half describing themselves as avid fans~\cite{harrypotterfan}. The book series sold over 600 million copies~\cite{harrypotterscholastic} and was translated into over 80 languages by 2020~\cite{TheRowlingLibrary-2020-BigTranslationsPotter-z}.

Set in modern England with magical elements like enchanted objects and spells, this context allows incorporating or discarding aspects of contemporary life, including social technology designs and constraints. Young protagonists and the coming-of-age narrative align with our focus on youth identity and relationship building. Character transitions parallel real-world moves between schools or locations, enabling exploration of how social technologies might support relationship maintenance across changing contexts. The series' emphasis on community and belonging suits examining youth identity expression and self-presentation in virtual spaces.

\subsection{Procedure}
The first author conducted all interviews over Zoom, with the second author joining one session to validate the FI method. Interviews averaged 90 minutes. We began by clarifying what 3D gaming experience means, asking participants what they like or dislike about 3D environments compared to 2D ones, and discussing \textit{Harry Potter} preferences (like favorite books) to break the ice and begin immersion in the Potter world.

We then presented the FI narrative: participants envisioned supporting social connection within Hogwarts as non-wizards (``Muggles''~\cite{muggle}) connecting across Hogwarts and Muggle worlds. We shared a Miro~\cite{miro} board~(Figure~\ref{fig:miro}) where participants sorted friends and family into Potter-inspired categories like houses (Gryffindor, Hufflepuff), teams (Quidditch members), and old friends (muggles). This encouraged thinking about relationships within the \textit{Harry Potter} context and enhanced fictional design space immersion.

During the design phase, we asked about social media experiences, focusing on meaningful versus meaningless aspects across existing platforms and what they envision in an ideal platform. While developing ideal magical social media capabilities and what life aspects they would share, we maintained narrative immersion by referring to existing platforms as ``Muggle social media'' and consistently engaging with the Hogwarts setting per FI guidelines. We also encouraged participants to think freely without the negative associations they might have with existing ``social media,'' asking questions like ``Imagine you're using this new magical platform---or maybe it's not even a platform. Whatever it is, what do you think this would look like?'' and ``If you could design this thing with any magical powers, what form would it take? (Is it a magical mushroom? A secret room? A creature?)'' We concluded by asking participants to reflect on how using their envisioned platform would feel and to compare it with their experiences on current social media. The semi-structured protocol is in Appendix \ref{protocol}.

\subsection{Participants and Recruitment}
We used purposive and convenience sampling. The first and second authors shared recruitment materials (including screener survey links) on social media, while the first author distributed campus flyers.

Participants needed to be 15--24 years old per UN~\cite{UnitedNationsUnitedNations-Other-Youth|Nations-t} and WHO~\cite{who-youth} youth definitions and consent to video/audio recording during 90-minute interviews. Video recording followed FI guidelines, enabling gesture analysis if needed. We included teenagers and young adults to capture perspectives from distinct life stages sharing focus on identity exploration, self-presentation, and relationships.

Because our FI sessions used a \textit{Harry Potter}-themed scenario to imagine digital spaces unconstrained by current design conventions, we required participants to have at least moderate familiarity with the franchise (agreeing with ``\textit{I remember some details about the plot, characters, and magical aspects}''). We did not require extensive knowledge, as deep familiarity with a specific fictional world seemed unnecessary for envisioning new possibilities for social connection and might introduce selection bias. We also required at least 20 hours of prior 3D gaming experience (e.g., \textit{Minecraft}, \textit{Roblox}). This criterion served as a baseline threshold for participants to draw on experiences with digital interfaces beyond 2D mobile screens when imagining alternative designs. While other modalities such as virtual reality could support such imagination, VR experience remains relatively uncommon among youth---only 24\% of teens have played games on a VR headset~\cite{Gottfried2024TeensGamesx}. In contrast, video gaming is nearly universal: 85\% of U.S. teens play video games, with the majority playing on consoles or desktop computers that render 3D environments~\cite{Gottfried2024TeensGamesx}. All but one participant met the gaming experience criterion. Although we imposed no constraints on social media experience, all participants reported some social media engagement.

\input{inserts/participants}
Eighty-four individuals completed the screener survey. We contacted 48 who met criteria; 23 provided consent and completed interviews. Mean age was 18.4 years (min=16, max=23, SD=1.56), with 15 (65.2\%) women, 3 (13.0\%) men, 2 (8.7\%) non-binary, and 3 (13.0\%) gender fluid participants. While our study is not intended to establish generalizable findings and our method was not designed with that goal~\cite{Daniel2019UsingResearchs}, we note that our participant demographics skew toward women. Full demographic information appears in Table \ref{tab:demographics-combined}.

\subsection{Data Analysis}
We analyzed interview transcripts using reflexive thematic analysis~\cite{BraunClarke-2021-ThematicAnalysisGuide-k}, which provides systematic yet adaptable qualitative data interpretation. Its independence from pre-existing theoretical frameworks suited our exploratory focus on youth-envisioned social media design through FI, enabling effective capture and analysis of diverse participant perspectives.

Since not all participants agreed to have their videos on during interviews, we did not conduct gesture analysis. Additionally, drawing was not part of our protocol because we have found in prior interviews that asking participants to draw can create high pressure for teen and youth participants, potentially limiting their imagination. All data therefore consisted of verbal interview transcripts.

Analysis began with the first two authors independently coding the same two transcripts line-by-line using descriptive codes closely adhering to data. Authors discussed and resolved coding discrepancies, then individually coded two different transcripts using Atlas.ti~\cite{atlas}. After four iterations, we reached code saturation. The first author then coded all transcripts using the final code set. All code versions and the final codebook are in supplementary materials.

\subsection{Ethical Considerations}
Given J.K. Rowling's controversial transgender statements~\cite{Other-2020-JKIssues-x, Miller-2024-JKTransantagonism-a}, we included a disclaimer in recruitment materials and surveys, reiterating our position during interviews. The disclaimer read: \textit{We acknowledge that J.K. Rowling has made statements regarding transgender individuals that many find harmful and discriminatory. We do not support or endorse these views. Our study focuses on the magical context of the Harry Potter series, and we will strive to create an inclusive and respectful environment for all participants, regardless of gender identity or expression. Thank you for your understanding and for helping us advance this important research.}

%% file: inserts/participants.tex
\begingroup
\begin{spacing}{1.0}
\small
\begin{longtable}{@{}l l l l l p{3.5cm}@{}}
    \caption{Participant demographics and envisioned social media features. Abbreviations for Harry Potter familiarity: Expert (remembers every detail), Proficient (remembers most details), Familiar (remembers some details). Abbreviations for platforms: IG (Instagram), DC (Discord), SC (Snapchat), TT (TikTok), BR (BeReal), X (Twitter/X), RD (Reddit), FB (Facebook), TH (Threads), LI (LinkedIn), NP (Noplace). Rows prefixed with ``>'' denote envisioned social media features.} \\
    \toprule
    \textbf{PID} & \textbf{Gender} & \textbf{Age} & \textbf{Race/Ethnicity} & \textbf{HP} & \textbf{Platforms} \\
    \midrule
    \endfirsthead
    \toprule
    \textbf{PID} & \textbf{Gender} & \textbf{Age} & \textbf{Race/Ethnicity} & \textbf{HP} & \textbf{Platforms} \\
    \midrule
    \endhead
    \midrule
    \multicolumn{6}{r}{\textit{Continued on the next page}} \\
    \endfoot
    \bottomrule
    \endlastfoot
    
    P01 & Woman & 19 & Asian & Proficient & IG, DC, BR, RD \\
    \multicolumn{6}{@{}p{\textwidth}@{}}{> Personalized magical houses; Pensieve-inspired memory sharing; Interactive pet guide; Neighborhood \& spatial socialization; Dynamic and context-aware avatars; Front porch for public updates and casual interactions; Public areas for open interaction \& interest-based rooms; Area-based access control; Music \& mood reflection in rooms.} \\
    \midrule
    
    P02 & Woman & 20 & Asian & Expert & IG, DC, SC, TT, BR, X, RD, FB, TH, LI \\
    \multicolumn{6}{@{}p{\textwidth}@{}}{> Pensieve-inspired memory sharing; Themed rooms for open discussions; Personal vaults; Dual-identity portal; House-elf moderator; Themed personal spaces; Ephemeral, once-view pensieve posts; Interactive room creation with spells; Vault access tracking; Serendipitous interactions via walking.} \\
    \midrule
    
    P03 & Woman & 18 & Black & Proficient & IG, DC, TT, X, RD, TH, NP \\
    \multicolumn{6}{@{}p{\textwidth}@{}}{> Personalized magical homes; Interactive town square \& community voting; Floating neighborhoods; Portable portal mirror; Pensieve-inspired memory \& emotion sharing; VR-like navigation; Selective visibility \& privacy settings; Shopping district \& immersive commerce; Customizable avatars \& dynamic appearance.} \\
    \midrule
    
    P04 & Woman & 16 & Asian & Expert & IG, DC, SC, TT, BR, X, RD \\
    \multicolumn{6}{@{}p{\textwidth}@{}}{> Floating memory orbs; Personalized spaces; Enchanted books with different features; Ghostly DJ with personalized music based on moods; Glowing message threads that float towards recipients; Wisp notifications; Interactive exploration in a village-like landscape.} \\
    \midrule
    
    P05 & Woman & 18 & White & Proficient & IG, DC, BR, RD, FB, TH \\
    \multicolumn{6}{@{}p{\textwidth}@{}}{> Immersive memory bubbles; Personalized spaces; Handwritten air messages; Magical jars for real-time communication; Interactive social memory town; Secret rooms for privacy; 3D commenting system; House elf companion (AI assistant).} \\
    \midrule
    
    P06 & Non-binary & 18 & White & Familiar & IG, DC, SC, TT, RD, FB \\
    \multicolumn{6}{@{}p{\textwidth}@{}}{> Customizable personal spaces; Co-created group spaces; Animated polaroid-style letters; Teleportation-based navigation; Memory-building activities; Interactive bulletin boards \& photo albums; Privacy through magical barriers; Intentional visit to friends' spaces without algorithmic scrolling; Neighborhood \& spatial socialization.} \\
    \midrule
    
    P07 & Woman & 17 & Hispanic/White & Proficient & IG, DC, SC, TT, BR, X, FB \\
    \multicolumn{6}{@{}p{\textwidth}@{}}{> Holographic interface; Brain-connected communication; Immersive memory sharing; Customizable profiles; Multi-threaded conversations; Personalized content curation; Integrated multi-platform content; Custom social interaction modes.} \\
    \midrule
    
    P08 & Woman & 19 & White & Proficient & IG, DC, SC, RD, FB \\
    \multicolumn{6}{@{}p{\textwidth}@{}}{> Pensieve-inspired memory sharing; Animated interactive painting; Messaging through magical letters; Color-coded wand lights for notifications; VR-like social spaces; Past interaction-based communication permissions; Customizable group interactions; Marketplace for magical item trading.} \\
    \midrule
    
    P09 & Woman & 21 & Asian/White & Proficient & IG, DC, TT \\
    \multicolumn{6}{@{}p{\textwidth}@{}}{> Personalized rooms; Themed social spaces; Space-based control access, secret areas, \& hidden objects; Magical mailbox system; Intentional content consumption; Neighborhood \& spatial socialization; Customizable avatars; Room of Requirement-inspired group chats.} \\
    \midrule
    
    P10 & Non-binary & 17 & Hispanic/White & Proficient & IG, DC, SC, TT, BR, RD, TH \\
    \multicolumn{6}{@{}p{\textwidth}@{}}{> Personalized rooms \& houses; Magical mailbox; Social village with communal spaces; Enchanted messenger animals; Immersive activity view; Customizable avatars \& profiles; 3D art sharing; Stores integrated with real-world services; Space \& object-based access control.} \\
    \midrule
    
    P11 & Woman/Non-binary & 19 & Native American/White & Familiar & IG, DC, TT, X, RD, FB, TH \\
    \multicolumn{6}{@{}p{\textwidth}@{}}{> Personalized houses; Interactive 3D avatars; Spatial discovery \& connection; Magical keys to access specific areas.} \\
    \midrule
    
    P12 & Man & 19 & Asian & Familiar & IG, DC, SC, TT, X \\
    \multicolumn{6}{@{}p{\textwidth}@{}}{> Immersive 3D social spaces; AI-generated personal worlds; Collaborative project spaces; Pensieve-inspired memory sharing; Temporary, safe-sharing space for struggles and emotional reflections; Adaptive privacy \& anti-scam protection; Interest-based connection zones.} \\
    \midrule
    
    P13 & Woman & 17 & Asian & Familiar & IG, DC, SC, TT, BR, X, RD, FB, TH \\
    \multicolumn{6}{@{}p{\textwidth}@{}}{> Personalized houses; Mutual QR-based friend verification; Dynamic room-based privacy; Shuffling access mechanism; Selective private comments; Entertainment \& saved content integration; Immersive 3D movement, interaction, and environment-based socialization.} \\
    \midrule
    
    P14 & Woman & 18 & Hispanic & Expert & IG, DC, SC, TT, BR, X, FB \\
    \multicolumn{6}{@{}p{\textwidth}@{}}{> Hybrid 2D-3D interface; Customizable personal rooms; Multi-layered privacy control; Teen \& child protection zones; Pensieve-inspired memory sharing; Real-time guided walk through personalized content; Gaming \& group spaces; Authentic avatars.} \\
    \midrule
    
    P15 & Woman & 20 & Asian & Proficient & IG, DC, SC, BR, X, RD \\
    \multicolumn{6}{@{}p{\textwidth}@{}}{> 3D virtual world; Personalized homes \& avatars; Teleportation; Invisibility cloaks for privacy; Magical visual effects for messages; Interactive social spaces; Streak-based engagement; Integrated marketplace \& digital collectibles; House elf AI assistant.} \\
    \midrule
    
    P16 & Man & 18 & Asian & Familiar & IG, DC, SC, BR, X, RD \\
    \multicolumn{6}{@{}p{\textwidth}@{}}{> Personalized virtual homes; Party system \& close-knit groups; Authentic avatars; Communal spaces for social activities; Teleportation \& open-world navigation; Digital memory archiving; Dynamic virtual world customization; Space-based access control.} \\
    \midrule
    
    P17 & Woman/Non-binary & 17 & Hispanic/White & Expert & IG, DC, SC, TT, X, RD, TH \\
    \multicolumn{6}{@{}p{\textwidth}@{}}{> Personalized 3D avatars; Personalized virtual homes; Interactive public spaces; Spontaneous social interactions; Integrated chat \& calls; Dynamic friend spaces; Intentional social connections; Space-based access control.} \\
    \midrule
    
    P18 & Woman & 23 & Asian & Familiar & IG, DC, SC, TT, X, FB \\
    \multicolumn{6}{@{}p{\textwidth}@{}}{> Teleportation into shared experiences; Customizable avatars \& environments; Interactive social hubs; Personalized virtual rooms; Public virtual rooms; Holographic comments \& real-time reactions; Ad integration without intrusion.} \\
    \midrule
    
    P19 & GenderFluid/Trans Man & 19 & White & Expert & IG, DC, SC, TT, X, RD, FB \\
    \multicolumn{6}{@{}p{\textwidth}@{}}{> Personalized homes; Virtual social activities; Selective social expansion based on interests; Immersive environment exploration; Multi-tiered social zones.} \\
    \midrule
    
    P20 & Woman & 17 & Black & Proficient & IG, DC, SC, TT, BR, X, FB \\
    \multicolumn{6}{@{}p{\textwidth}@{}}{> Immersive, interactive holograms for music \& gaming; Anonymous discussion threads; Contextual commenting system; Interest-based content; Real-time online status indicators; Personalized comment moderation.} \\
    \midrule
    
    P21 & Woman & 17 & Asian & Expert & IG, DC, SC, RD \\
    \multicolumn{6}{@{}p{\textwidth}@{}}{> Emotion-revealing teleportation; Magic book messaging; Marauder's map social tracking; Personalized homes; Emotion-based avatars; Selective mind reading for close friends; Compatibility-based social matching; Spontaneous conversation via walking; Immersive social events; Space-based access control.} \\
    \midrule
    
    P22 & Woman & 19 & White & Familiar & IG, DC, SC, TT, X, RD \\
    \multicolumn{6}{@{}p{\textwidth}@{}}{> Personalized avatar; Personalized virtual houses; Public \& private houses; Town square \& bulletin board; Interest-based groups \& activities; Integrated messaging system; Personalized profile pages; Adult \& teen separation.} \\
    \midrule
    
    P23 & Man & 18 & Asian/Hispanic & Familiar & IG, DC, TT \\
    \multicolumn{6}{@{}p{\textwidth}@{}}{> Personalized homes; Neighborhood \& social spatialization; Animated sand avatars; Communication via sand figures; Real \& anonymous dual identity; Interest-based social discovery; Dynamic social spaces; Space-based access control; ``Hall of Fame'' memory posts.} \\
    
\end{longtable}
\end{spacing}
\endgroup
\label{tab:demographics-combined}


%% file: sections/4_results.tex
\section{Results}
Drawing from their experiences with 3D games and familiar media worlds, participants imagined social interactions extending beyond mainstream platforms like Instagram. Through the speculative frame of exploring a ``new magical `thing' for social communication'' for Hogwarts students, they envisioned how digital environments might enable more embodied, relational, and emotionally meaningful connection. Their visions converged around six interrelated themes that collectively contribute to a vision of social media that is more meaningful, authentic, and aligned with their social and emotional needs. We refer to this synthesized concept as ``Social Media at Hogwarts,'' or SMH, throughout this paper.

In line with the goals of co-design processes~\cite{Sanders2008Co-creationDesignh}, our interest lay \textit{less in the specifics of functionality and more in uncovering what kinds of interactions and environments participants valued most.}

\subsection{Theme 1: Intuitive Social Navigation}
\label{section:4-1}
Youth sought interactions that feel natural, embodied, and socially meaningful—reflecting how people navigate relationships and spaces in everyday life. They imagined SMH not as a collection of buttons, but as a space where presence, movement, and context shape connection, self-expression, and privacy in intuitive, familiar ways.

\subsubsection{Supporting Presence and Connection Through Embodied, Immersive Interaction}
\label{section:4-1-1}
Embodied, immersive qualities—realistic movement, lifelike avatars, holographic interfaces—could transform SMH into a space fostering presence and emotional connection. Current social media (``Muggle Social Media'' or MSM) was described as doing \inlinequote[16]{a good job at spreading information [but not] emotions.} P01 captured this limitation: \inlinequote{if you're just watching things through a screen\ldots{} technically, there's like a wall dividing you.}

For others, fulfillment came from embodied physical actions. P04 valued \inlinequote{meandering through different people's spaces,} finding deliberate exploration heightened their sense of being \inlinequote{more present in somebody's space.} P09 appreciated how moving physically between rooms created a \inlinequote{feeling of time passing}: \inlinequote{there's a feeling like you're actually officially leaving a space as opposed to clicking into a different one.} Traversal in a \inlinequote{corporal form} would feel \inlinequote{fulfilling,} \inlinequote{calming,} and \inlinequote{reassuring.}

P03 reflected that simple actions like \inlinequote{mov[ing] your legs to walk} or \inlinequote{turn[ing] your head} add a \inlinequote{tiny bit of connection,} making it \inlinequote{easier to have connections with people.}

This sense of presence emerged even in seemingly passive actions. As P01 explained,
\blockquote{I don't know, there's just a feeling about it that makes it different. I can't really explain. Depending on 2D and 3D, but I feel like in 2D, you can't really change anything about their homepage. You just observe it. The most you can do is add a like or something, and that would change their like count. But then they would know who you are, so that takes away the anonymity. But if you're just looking at it, then you're not really affecting anything. That's just for your knowledge. But in a 3D setting, even if you don't leave any physical footprints, in a way, they're still there because you opened the door, you went in, you sat down. Maybe just that mark in a way that you've left on the place takes away from the anonymity. I don't know if that makes sense but.}{01}
For P01, 3D interaction introduced a felt presence that lingered, even without direct interaction.

\subsubsection{Using Spatial Cues and Social Norms to Manage Access and Privacy}
\label{section:4-1-2}
Participants proposed spatial and visual cues modeled after real-world environments to make access and privacy management feel more intuitive and contextually appropriate. P03 described public areas like \inlinequote{the downstairs area,} \inlinequote{guest bedroom,} or \inlinequote{game room} as spaces where people could \inlinequote{just come in} without permission, while private spaces like their bedroom required \inlinequote{extra consent} as it was a \inlinequote{safe space.} P13 suggested intermediate zones like a \inlinequote{waiting room} for sharing \inlinequote{semi-personal things} before granting further access.

P01 contrasted 2D and 3D settings, noting how boundaries over personal space are clearer: \inlinequote{In 2D, you're just observing.\ldots{} But in a 3D setting, even if you don't leave physical footprints, in a way, they're still there.} They expected others to recognize and respect spatially cued privacy norms:
\blockquote{I don't think people should be able to request access to areas when you're not there.\ldots{} Why would you want to go into my virtual bedroom when I'm not even home? If I wouldn't do it in real life, I probably wouldn't want to do it online either.}{22}

While managing visibility across multiple rooms might seem complex, participants believed this approach would feel intuitive. P11 suggested spatial contexts could reduce privacy control efforts: \inlinequote{It would make things easier because you'd be able to easily tell who's allowed and not allowed in certain areas of the house.}

\subsection{Theme 2: Shared Experiences}
\label{section:4-2}
Shared experiences—through active collaboration or simply being together—play a crucial role in forming meaningful relationships and lasting memories.

\subsubsection{Fostering Shared Memories Through Collaborative Activities}
\label{section:4-2-1}
Shared activities served as meaningful foundations for building friendships. P06 contrasted casual solo games with more immersive, co-experienced ones. While playing a simple emoji game lacked true shared experience since \inlinequote{we don't actually see each other's screens,} Roblox fostered lasting memories: \inlinequote{do you remember that time when that kid in the voice chat said this, or when we got scared by that jump scare... it's not just that we're in the same space, but that we're doing something together and experiencing it together.}

P12 described the value of collaborative projects like \inlinequote{creat[ing] a house we love} online, which offered both \inlinequote{a sense of self-achievement} and \inlinequote{teamwork,} contrasting with the passivity of \inlinequote{just like doom scrolling.} Working together on \inlinequote{the same topic, on the same project} felt like \inlinequote{a better way to connect} than \inlinequote{posting} content individually.

The value of shared experiences extended beyond completing tasks. P21 described how walking could spark memorable conversations: \inlinequote{If I'm outside and we're all in the mood to walk, then we can just choose to walk to the cafe\ldots{} Because many times when you're walking together, you come up with the most out-of-pocket conversations, and sometimes the best memories happen during a walk.}

\subsubsection{Making Space for Casual Connections Through Low-Pressure, Ambient Social Interaction}
\label{section:4-2-2}
Low-pressure, ambient interaction—co-existing in shared spaces or listening to music together—offered natural ways to build connections. P18 described a \inlinequote{study sesh} where friends could send avatars to \inlinequote{sit and study together} in a virtual room with Lo-Fi music, mirroring Discord servers' 24/7 activity and offering comforting presence without direct conversation. Even \inlinequote{mindless scrolling} became meaningful when shared: \inlinequote{see[ing] their reaction, hav[ing] a discussion, and\ldots{}commentat[ing] in real-time, instead of just sending it back and forth,} would feel \inlinequote{10 times more fulfilling.}

Music often facilitated these interactions. P19 envisioned a \inlinequote{jukebox} where friends could \inlinequote{queue up songs}:
\blockquote{At one point, literally, we sat for an hour in silence, listening to a whole playlist\ldots{} we were just sitting there listening, vibing the music. And it was one of the most awesome nights because we all knew in that moment that collectively we all enjoyed what we were hearing and it was just good. I don't know how else to describe it. It was just nice.}{19}
Music served as both \inlinequote{conversation starter} and passive connection: \inlinequote{you don't have to connect through talking to someone about music. You can just listen to the music.}

\subsection{Theme 3: Communal Ambience}
\label{section:4-3}
SMH was envisioned as a platform for intentional, engaging interactions with close friends, making it naturally less prone to social comparison. P09 described it as \inlinequote{the antonym to pressure,} \inlinequote{stress-free,} \inlinequote{community oriented,} \inlinequote{wholesome,} \inlinequote{accessible,} and \inlinequote{accepting.} Unlike MSM, which often feels performative and stressful, SMH would foster organic, personalized interactions and closer friendships through neighborhood structures, shared gathering spaces, and stronger boundary controls.

\subsubsection{Designing Neighborhood Structures that Reflect Real-World Communities}
\label{section:4-3-1}
SMH's 3D world would mirror real-world communities, using familiar layouts like homes, neighborhoods, and shared areas to foster interaction. P22 described this in detail:
\blockquote{So, in a virtual world, you might have your house, and then you could go to a community board or town square with postings for different groups\ldots{} From there, you could travel to those group spaces. I don't know if you've ever played or seen Club Penguin, but in that world, you had your igloo to decorate and could visit the town\ldots{} The town had different buildings for different activities. I'm imagining something like that, where you have your house where people can visit and talk with you, and if you want to meet new people, you can leave your house and go out into the town.}{22}

Physical distances within the landscape would represent social tie strength. One participant elaborated:
\blockquote{And in this magical world, I imagine the people around you in your neighborhood would be from the same house as you, kind of like how people live together in dorms on a college campus. The nearby stores would be the ones your group usually goes to. So you'd share that sense of togetherness because you all go to the same stores or belong to the same house. I think that would help build a real sense of community.}{03}

P01 proposed a tiered layout where \inlinequote{the top 7 people you're closest to would be the 7 houses in your cul-de-sac area,} while acquaintances would be farther away, requiring a \inlinequote{little hike} to reach. This alignment of physical landscape with emotional closeness would streamline interactions. As P07 noted, in short bursts of social media use, they preferred seeing content from their closest friends over \inlinequote{some girl I talked to once three years ago.}

\subsubsection{Establishing Third Places for Casual Social Interaction and Serendipitous Connections}
\label{section:4-3-2}
SMH neighborhoods would include shared third places where people could casually gather, lowering social barriers and creating connection opportunities. P05 described these as \inlinequote{neutral spaces} not tied to specific activities or relationships, with \inlinequote{some places to sit, and some kind of flat table-like surface} supporting casual activities like \inlinequote{doing homework or study, or [playing] some kind of board game.} Such spaces would function as \inlinequote{a third, neutral territory} for organic interaction without managing invitations or switching platforms.

Shared neighborhood spaces—common rooms, courtyards, clubhouses—would allow casual interaction outside personal spaces. P01 described gathering in a \inlinequote{courtyard or game room} at the neighborhood center. Drawing on Club Penguin~\cite{club-penguin} and Poptropica~\cite{FeigenbaumOtherPoptropicaGamesb}, they imagined a \inlinequote{clubhouse\ldots{} open to anybody,} with themed rooms catering to different interests, like a Quidditch field or library, creating opportunities to meet others through shared activities. P09 imagined a \inlinequote{virtual dining room} for casual conversation while eating—open and inclusive, enabling cross-social group connections unlike Hogwarts' segmented house-based accommodation.

Public third places would also support serendipitous interactions. P03 described chance encounters while shopping: \inlinequote{you strike a conversation because you're both picking up the same shirt\ldots{} little connections like that happen just because you're in a public space.} P21 imagined \inlinequote{small arenas} for \inlinequote{Quidditch games\ldots{}[where users] can choose to get seated randomly [and] just start talking to the person next to you.}

\subsubsection{Prioritizing Close Relationships and Reducing Extraneous Content}
\label{section:4-3-3}
To preserve SMH's focus on meaningful, stress-free interactions, participants emphasized filtering irrelevant content like celebrity drama or political commentary while prioritizing updates from close friends. P19 criticized how platforms like Instagram have shifted away from their initial focus:
\blockquote{We need old Instagram back in 2010, when my mom was posting me and my brother just for the family and no one else. We didn't need the explore page. We didn't need Instagram shopping\ldots{} We don't need extra things in an app that is only made for one thing---connecting\ldots{} We need to figure out how to put boundaries in place between connecting with the people we love and connecting with people we don't know yet\ldots{} you don't need to have so many political takers on this one app\ldots{} It's just a very stressful situation.}{19}

Several participants advocated minimizing celebrity influence, with P01 suggesting isolating them to a \inlinequote{celebrity zone.} P07 captured the preference: \inlinequote{If I'm just gonna open the app for like 5 seconds\ldots{} I'm gonna wanna see my best friend's post, not some girl I talked to once 3 years ago.} SMH would be \inlinequote{more designed for people you actually know\ldots{} like your friends, classmates, or Quidditch team.}

\subsubsection{Developing Socially Grounded and Themed Conversational Spaces}
\label{section:4-3-4}
Themed, socially grounded conversational spaces for protected, meaningful dialogue could make SMH feel more personal than MSM.
\blockquote{Anyone you're talking to in general, just emphasizing conversation between individuals more, rather than just posting on a thread. Like how I talked last time about Reddit, you know, there's a thread, and then you make a comment on it, but there are a hundred other comments, so no one's going to respond to you. I want to focus more on having actual conversations with people, making real connections, and more individualized chatting. I find it's way easier to get to know people a lot better.}{22}

Many valued walking into themed, topic-specific rooms where conversation felt protected and purposeful, rather than broadcasting to broad, undefined audiences. P22 imagined \inlinequote{chat rooms you can walk into,} each centered on themes like \inlinequote{high school stuff,} decorated with artifacts from shared experiences. These would be \inlinequote{protected space[s]\ldots{}where you can talk\ldots{}without feeling unsafe or judged,} contrasting with MSM's more exposed feel where \inlinequote{you don't know everything around it.}

Personalizing these spaces would make them more emotionally connected. P07 described filling chatrooms with \inlinequote{photos of you together, inside jokes, or fun memories.} P09 imagined a \inlinequote{Room of Requirement}-like system~\cite{room-of-requirement} where users could create custom rooms on demand to match the \inlinequote{purpose} of their conversation. Interest-based spaces, similar to Discord~\cite{discord}, would provide a \inlinequote[13]{common point} or \inlinequote[07]{good talking point} to break the ice.

\subsubsection{Implementing Comprehensive Boundary Controls to Create Personal Spaces that Feel Safe}
\label{section:4-3-5}
SMH would need comprehensive boundary controls beyond existing platforms' limited and inconsistent privacy tools. P22 explained that even with contact syncing off, Instagram still suggested their Finsta (secondary account~\cite{VitakHuang-2022-FinstaAccounts-t}) to people they wanted to avoid, based on mutual follows. They stressed needing stronger blocking controls, including blocking \inlinequote{alternate accounts} and fully removing blocked person traces: \inlinequote{I don't wanna be able to tell that the person I blocked is even around me.} P14 pointed out that even private Instagram accounts expose personal content like Story Highlights to \inlinequote{a random person}, revealing perceived privacy protection gaps~\cite{story-highlights}.

Participants wanted stronger control over visibility, imagining features letting them opt out of being seen entirely. P03 suggested an \inlinequote{invisibility cloak} for moments when users \inlinequote{don't want anyone to see you.}

Preventing unsolicited stranger contact was another priority. P22 proposed a Discord-inspired model where only people you've interacted with can contact you, while P13 argued people \inlinequote{should only be able to add you if you meet in person and exchange a QR code,} reflecting that \inlinequote{it's too easy to find people on social media.}

\subsection{Theme 4: Nuanced Self-Presentation}
\label{section:4-4}
Participants described MSM as \inlinequote[22]{a popularity contest} that felt \inlinequote[14]{really fake} because \inlinequote[14]{people only really show the good moments}. This curated image culture creates \inlinequote[23]{a lot of seriousness}, reflected in how \inlinequote[23]{actors or celebrities\ldots{} [even] have somebody else run their Instagram account}. P13 shared helping a friend prepare an Instagram post, including selecting \inlinequote{post order,} highlighting \inlinequote{how seriously we take social media and our image online.} SMH, by contrast, would allow self-presentation that feels more personal, authentic, and flexible—reflecting users' full range without curation or performance pressures.

\subsubsection{Expressing Identity Through Customizable Avatars, Spaces, and Artifacts}
\label{section:4-4-1}

Customizable avatars and personal spaces offered meaningful self-expression far beyond MSM's limited settings. P21 envisioned building their own \inlinequote{mini world} complete with \inlinequote{houses, teleport points, and rooms like a kitchen, a bathroom, a bedroom, or even a school,} sharing these spaces as \inlinequote{showing a little part of yourself.} P03 wanted to create a \inlinequote{dark academia} Hogwarts-inspired room, filled with \inlinequote{so many bookshelves and\ldots{} Victorian architecture.}

Customizable \inlinequote[11]{3D avatar[s]} and profiles were key ways of showing individuality. P22 reflected on early Myspace~\cite{myspace}, describing how they enjoyed coding profile layouts and changing \inlinequote{the font and really make it reflect me.} Even small details like choosing usernames could \inlinequote{convey to the world and your friends your own identity\ldots{} which is a positive feeling usually.} P01 noted such expression is often \inlinequote{mainly for you,} not just for public display. P17 added that even unconscious choices reveal something personal: \inlinequote{your brain conjured that up, right? So I think that's part of you.}

Beyond visual customization, SMH would support ambient, indirect expression—subtle cues about what someone is doing, feeling, or thinking without formal posts. P01 wanted their avatar to reflect real-time states like current outfit or music, so friends could intuitively notice \inlinequote{`Oh. that's what she's listening to right now' or `That's what she's thinking about.'} P10 imagined friends picking up on interest changes by observing virtual space updates over time, providing \inlinequote{an insight into\ldots{} what I'm thinking.}

The contrast between 2D platforms' flat feel and 3D environments' immersive qualities mattered. P16 noted that VR creates presence feeling more like meeting \inlinequote{real people\ldots{} compared to profile pictures,} because you can pick up on \inlinequote{mannerisms} and details like \inlinequote{the way they hold themselves\ldots{} the way they decorate their house\ldots{} how they choose to dress their in-game character.}

These choices feel more authentic than generic posts because they emerge naturally from personal preferences. As P01 put it, such \inlinequote{creative expression} feels \inlinequote{a lot more personal} because \inlinequote{that's not just like what life has given you\ldots{} a reflection of your outlook on life.} P13 explained that these subtle forms of sharing make social media feel more like \inlinequote{an extension of everyday interaction,} helping people \inlinequote{gain the basic foremost information} about someone without direct conversation. P21 suggested this would make it \inlinequote{easier for people to approach you and say, `Hey, I like that, too.'}

Personal spaces like bedrooms carry intimacy and authenticity that translates well to SMH. P13 explained:
\blockquote{A bedroom is a respected personal space\ldots{} If you label it as a bedroom, it just feels more intimate. It's where you spend most of your time, and it shows your personality the most. It's not like you can easily change the stuff you have in your room.\ldots{} If someone were to come over, you can't just rearrange it in 30 minutes or 30 seconds the way you can on Instagram. If you wanted to add someone there, all you'd have to do is remove a couple of things. It's easier than physically changing a space.}{13}

These ambient, low-pressure forms of expression also carried lower social stakes. P16, who described themselves as \inlinequote{fairly private} and not particularly \inlinequote{public} about their preferences, explained that \inlinequote{decorat[ing] a house} or customizing an avatar allows others to \inlinequote{get to know more about my personality\ldots{} through my style} without feeling performative. This kind of expression felt \inlinequote{more justified} and less \inlinequote{embarrassing} than posting content just to be seen.

\subsubsection{Enabling Multifaceted Self-Presentation Across Spaces and Relationships}
\label{section:4-4-2}
SMH would support flexible, multifaceted self-presentation, making it easier to share different sides across various spaces and relationships without pressure to maintain a single, polished identity. P01 described having a house with rooms \inlinequote{dedicated to each of your interests}: \inlinequote{I'm into multiple things. You could have a room for Harry Potter, a room for neuroscience, and a room for another fandom or interest.} Representing different facets would reduce misjudgment and give others a more complete picture. P02 imagined users with multiple \inlinequote{Muggle personalities} organizing them into separate spaces: \inlinequote{Each personality has its own vault.}

\subsubsection{Sharing Everyday Activities and Lived Experiences to Deepen Understanding and Connection}
\label{section:4-4-3}
Participants valued sharing casual, everyday moments and rich, embodied memories—filling gaps between MSM's typical highlights to create a more natural extension of daily life. P04 wanted to share small, ordinary moments, like \inlinequote{what I'm doing at the moment. For example, if I'm at the shops, maybe I'd want to share that.} P21 extended this to items purchased, such as a \inlinequote{new pair of New Balance shoes,} imagining virtual spaces that automatically displayed \inlinequote{recently added or recently stored items.}

SMH could also support personal, embodied memory-sharing to deepen understanding and reduce misinterpretation. P04 wanted to share lived experiences like \inlinequote{jumping into a pool or taking the first bite of an ice cream sundae,} drawing inspiration from the \textit{Pensieve} in Harry Potter~\cite{pensieve}. Unlike MSM, where communication often feels like \inlinequote[03]{just words,} memory-sharing could communicate context, emotion, and intention more fully. P02 explained this could help others \inlinequote{understand why I said what I said} and prevent \inlinequote[02]{misconstru[al]s.}

Some imagined shared memories including sensory or emotional details beyond visuals. P01 expressed interest in being able to \inlinequote{transmit smells through social media.} Others compared it to live streaming personal experiences—P07 described it as \inlinequote{screen sharing, but with your eyes,} and P08 highlighted its usefulness for capturing fleeting moments, like crossing a state line, without pausing to record. P03 imagined instantly sharing moments with loved ones \inlinequote{in the moment,} rather than explaining them after the fact when the emotion has passed.

\subsubsection{Flexible, Contextual Privacy as a Foundation for Authentic Self-Presentation}
\label{section:4-4-4}
Flexible, contextual privacy is essential for authentic self-presentation—allowing people to manage who sees what, when, and how without feeling overexposed or misrepresented. Current platforms offer rigid, limited controls that fail to account for real social life's complexity. P16 explained that on Instagram, privacy options are reduced to features like \inlinequote{close friends}, adding, \inlinequote{it's kind of a pain to organize.} P11 noted that as people share \inlinequote{more like their genuine real life and their experiences,} privacy needs become \inlinequote{a little bit more complicated,} requiring finer-grained visibility control.

SMH could support more seamless transitions between identities and spaces. P02 described a system that would \inlinequote{just telepathically know what's in Muggle area and what's in Hogwarts area,} allowing users to \inlinequote{walk through a portal} and appear as \inlinequote{a new person} without needing separate accounts. Users could shift how they present themselves based on audience or situation. P03 proposed a model where everyone maintains a \inlinequote{base} identity but customizes additional layers. P22 described having different avatars for different groups, like on Discord, where \inlinequote{any group you're in\ldots{} you could change your avatar for that.}

Contextual privacy would extend beyond people-based access to include temporal and relational factors. P14 proposed limiting access to different stages of one's past identity: \inlinequote{I'm cool with you seeing the 6-year-old version of me on this, but I don't want you to see my 12-year-old version.} P21 suggested gradual openness, where people could selectively reveal more over time: \inlinequote{if you're meeting a complete stranger, you wouldn't want to give them like exactly how you look\ldots{} But then, after a couple of conversations, you sort of get the gist of how they are, and whether you want them to see you\ldots{} so then you can turn it on. Or turn it off.}

Individual control over these transitions mattered. P14 explained that privacy settings should not be one-size-fits-all: \inlinequote{each of them should have their own setting, because I feel like everybody has different levels of what they want to keep to themselves.} Together, these suggestions reflect flexible, contextual privacy not just as a technical feature, but as a core requirement for the personal, meaningful interactions youth hoped SMH would enable.

\subsection{Theme 5: Intentional Engagement}
\label{section:4-5}
Participants described growing frustration with current social media prioritizing passive consumption over meaningful connection. SMH would promote intentional, socially rewarding interactions—refocusing social media on authentic connection, purposeful engagement, and meaningful time use.

\subsubsection{Encouraging Deliberate Navigation and Content Discovery}
\label{section:4-5-1}
SMH would encourage deliberate, effortful discovery, replacing MSM's passive, algorithm-driven feeds with experiences requiring users to actively choose what and who they engage with. P22 described frustration with Instagram: \inlinequote{half of the timeline is accounts I don't even follow that were just randomly recommended to me,} turning it into \inlinequote{less keeping up with friends and more just getting random stuff shoved in my face.}

P04 described how slowing down to explore someone's space or story could make interactions feel more personal. Unlike Instagram, where you \inlinequote{see all the posts at once} and can quickly scroll past them, SMH would require users to \inlinequote{take more time to hone in on a memory,} offering the satisfaction of consciously consuming content from someone you're close to, rather than \inlinequote{taking a quick second to look at it\ldots{} and then being done.}

Some imagined features prompting users to actively choose what they wanted before entering a space. P09 compared this to using an elevator, where you decide on your destination in advance: \inlinequote{Could [SMH] be sort of like that, where you think about what you want to see before entering that space, and then it shows what you were thinking about?} Others imagined brain-controlled interfaces surfacing content based on what users genuinely want in the moment, not just what an algorithm predicts will keep them engaged.

Physical movement through SMH's world was key to this intentionality. P03 explained that embodied movement makes SMH feel \inlinequote{more connected than regular social media} because it requires active participation rather than passive scrolling. P02 preferred slower, more thoughtful interactions over instant teleportation: \inlinequote{I wouldn't really want anything to be super fast decisions. I want everything to be very thought through.}

P05 imagined SMH as a world where you might \inlinequote{walk through a town to get there}: \inlinequote{It's not challenging to navigate, but you have to intentionally go there to see it. It isn't something you can do mindlessly.} P14 agreed that visiting someone's space in SMH would feel much more personal than encountering \inlinequote{a random 2D Insta post that they posted 3 months ago.}

This effortful navigation would naturally discourage superficial engagement. P05 noted: \inlinequote{Influencers probably wouldn't be as keen on something that is more time-consuming\ldots{} People who are genuinely just trying to connect with others would probably benefit\ldots{} because it is more personal.}

\subsubsection{Making Social Media About (Meaningful) Connection, Not (Passive) Consumption}
\label{section:4-5-2}
Beyond content navigation, participants emphasized shifting the platform's core purpose from endless consumption toward meaningful social connection. They expressed frustration with how easily time slips away on platforms like Instagram and TikTok, leaving them unfulfilled. P14 captured this, describing time spent on Instagram Reels while \inlinequote{literally just laying in [their] bed} as \inlinequote{wasting [their] time.}

MSM's infinite scrolling design contributed to this feeling. P11 explained \inlinequote{there's always more content on social media,} fueling fear of missing out and urges to keep scrolling. P14 critiqued TikTok for exposing users to \inlinequote{clickbaity} low-quality content that \inlinequote{doesn't stimulate [users] at all} and \inlinequote{messes with [their] attention span.}

Active, social engagement felt far more rewarding. P16 shared that \inlinequote{anytime not scrolling} but \inlinequote{talking to people} feels worthwhile, \inlinequote{even if it's just light conversation.} P14 noted meaningful use comes when \inlinequote{actually talking to someone,} especially staying connected with people \inlinequote{not in close proximity.} They reflected that while users might spend more time on SMH, it would feel worthwhile because \inlinequote{at the end of the day, you're connecting with a person that you can't otherwise connect with\ldots{} and that makes it more valuable in and of itself} compared to watching \inlinequote{50-second TikTok videos of somebody making slime for like an hour.}

Participants contrasted passive consumption with active participation in interactive environments. P10 explained that while doomscrolling feels unproductive, activities like Minecraft offer \inlinequote{progress} that can be \inlinequote{saved and come back to.} P06 noted they \inlinequote{don't feel bad\ldots{} most of the time [they] use Roblox,} because it involves \inlinequote{playing games with friends} and \inlinequote{customizing spaces.}

P09 acknowledged \inlinequote{time spent might not be mitigated,} but emphasized \inlinequote{the way you're spending it} would make the difference. SMH would \inlinequote{feel a lot more interactive instead of just taking in content, and more and more and more content}—meaningful engagement, rather than passive consumption, ultimately mattered.

\subsection{Theme 6: Playfulness and Whimsy}
\label{section:4-6}
Participants described SMH as a platform feeling youthful, playful, and deeply immersive, offering entertainment plus opportunities for creative self-expression, social bonding, and emotional well-being. By embracing whimsy and delight, SMH would make social media feel more personal, less pressured, and more meaningful.

\subsubsection{Embracing Creative Joy and Playful Self-Expression}
\label{section:4-6-1}
Participants imagined SMH transforming social media into a space for creative joy and playful self-expression, offering freedom to experiment with identity, explore personal interests, and connect through creativity. P11 imagined a \inlinequote{dream social media world} where users could fully redesign their profiles, \inlinequote{maybe their avatar in the middle, maybe their posts on either side, maybe a bio below}, allowing them to \inlinequote{express their creativity and, like, yourself} in new ways not possible on conventional platforms.

For some, personal customization felt like meaningful identity exploration. P01 described the deeper satisfaction in \inlinequote{just that process of creating it, and finding out who you are through that process.} P21 echoed this sentiment, describing SMH as feeling \inlinequote{more like a game} because of the \inlinequote{creative freedom} it provides, making the experience feel \inlinequote{more fictional than real} in a way that encourages experimentation and self-expression. P22 added that SMH offered something they hadn't found anywhere else, explaining, \inlinequote{I can really be creative here\ldots{} I can make something that I haven't made anywhere else, you know. That's where the fun aspect comes in, really, for me.}

This creativity extends beyond individuals to the wider environment and community. Participants described the excitement of exploring what others had created, likening it to visiting realms in Minecraft where you \inlinequote{get to see how like everyone puts in their own effort into it\ldots{} Whoa! They built that, that's so cool} (P06). SMH felt alive and engaging whether users were socializing or spending time alone: \inlinequote{it's not just like, Oh, the group chats are only fun when there's other people here. No, it's fun when we're by ourselves, too, we get to do our own things} (P06). The interactive environment itself, where \inlinequote{you're not just sitting there chatting, you're also walking around the space\ldots{} you're interacting with whatever else is in that area} (P06), offered more to do than conventional platforms. Whether through building, exploring, or engaging with shared content, SMH transformed social interaction into something playful, creative, and deeply engaging.

\subsubsection{Designing Engaging and Playful Privacy Mechanisms}
\label{section:4-6-2}
Participants imagined privacy not as tedious and confusing but as part of playful experience, envisioning interactive, game-like mechanisms turning boundary-setting into something fun, discoverable, and socially meaningful. P11 envisioned controlling space access through gestures like \inlinequote{secret knock} or hidden objects revealing \inlinequote{secret passages} for trusted friends. Others suggested knowledge-based gates like \inlinequote[21]{secret handshakes} shared among friends. Drawing from the Marauder's Map, P02 imagined privacy controls activated by \inlinequote{spells}, unlocking spaces when correctly summoned.

P05 compared these features to \inlinequote{little Easter eggs,} remarking \inlinequote{there's that little kid part of everyone} that makes such features enjoyable, contrasting them with mainstream social media's \inlinequote{wall of settings} that often feel confusing.

\subsubsection{Augmenting Everyday Interactions with Fantastical Interfaces and Environments}
\label{section:4-6-3}
Participants imagined SMH using magical and fantastical technologies to transform everyday digital interactions, making them feel more immersive, accessible, and emotionally uplifting, even for those with limited real-world opportunities. P01 questioned the need for phones at all, envisioning wands projecting \inlinequote{holographic screens, like Jarvis in Iron Man,} or activating interfaces through spells. Participants also imagined alternative magical devices: VR glasses, a \inlinequote{makeup mirror} portal, or a \inlinequote{watch} that immerses users into SMH.

These magical affordances were not only imagined as novel interfaces but also as ways to make extraordinary experiences more accessible. P09 highlighted that SMH could give people \inlinequote{experiencing stress} or lacking access to \inlinequote{really, really beautiful landscape and environments} the chance to walk around and \inlinequote{admir[e]} calming, scenic spaces. P16 imagined that users could take part in activities they might never otherwise experience, saying, \inlinequote{Maybe you're busy\ldots{} you can't afford to travel to Nepal to climb Mount Everest\ldots{} but you have a day, so you can just log in for the entire day and just climb a mountain.}

\subsection{Synthesis: SMH as a Hopeful Reframing of Social Media}
\label{section:4-7}
Given the speculative FI co-design workshop, SMH embodies youth aspirations and values for social media platform interactions. While acknowledging potential challenges like over-immersion or distraction, they viewed these downsides as worthwhile trade-offs for its benefits. Participants described SMH as an inspiring alternative to existing social media, offering self-expression opportunities and deeply personal, intentional connection. They saw SMH not just as a better platform, but as an inspiring redefinition of social media's purpose and potential.

\subsubsection{Breaking Free from Platform Monotony with Novel Interactions}
\label{section:4-7-1}
Participants saw SMH as offering better user experiences, departing from existing platforms' repetitive structures, and inspiring hope for social media's future. P06 appreciated how SMH combines desirable aspects: \inlinequote{You get friends who you can talk to, keep up with, and stay connected to. And you get things you can customize. You kind of get the whole package there.}

Participants appreciated SMH's break from existing platform monotony. P11 pointed out that most social media platforms share a common \inlinequote{standard} with repetitive features like \inlinequote{feed, messages, settings,\ldots{} profile.} In contrast, SMH introduced new types of interactions that deviated from familiar patterns. P02 described SMH as \inlinequote{a combination of Discord and BeReal} that brings together casual sharing and topic-based discussions. P17 envisioned SMH as \inlinequote{more interactive, kind of like if Discord and Instagram were combined,} enjoying Discord's interactions with Instagram's profiles and posting ability. P22 highlighted limited customization on other platforms, explaining \inlinequote{I just have a profile picture.} On SMH, they could \inlinequote{really be creative} and \inlinequote{make something that [they] haven't made anywhere else,} making the experience feel \inlinequote{novel,} like a \inlinequote{breath of fresh air.}

For some, SMH's novelty lay in prioritizing enjoyment over seriousness. P14 described initially defaulting to the idea of a more ``mature'' 2D version, but later embracing the playful, immersive 3D version as more personally meaningful and fun. For example, P14 initially preferred a 2D version but shifted to 3D, reasoning: \blockquote{Yeah, [I would prefer the 3D version] actually 100\%. I feel like,\ldots{} the only reason I said 2D is because I felt like the need to be like a mature adult. But who cares? I feel like something that's more enjoyable would be [the SMH].}{14} P22 described SMH as \inlinequote{the coolest,} \inlinequote{childlike,} \inlinequote{less generic,} and \inlinequote{a lot more fun} compared to other platforms. P06 observed that \inlinequote{the physical space} on SMH felt \inlinequote{a lot more fantastical\ldots{} whimsical, fun} compared to \inlinequote{traditional ones.} The platform also evokes nostalgia and happiness. P15 remarked \inlinequote{Having this app would actually make them feel happy\ldots{} with nostalgia,} anticipating SMH would be \inlinequote{a super big hit} with teenagers due to its playful, engaging nature.

\subsubsection{Inspiring Hope and Expanding the Scope of Social Media Interactions}
\label{section:4-7-2}
Through the FI process, participants realized they weren't limited by existing platform structures. Instead, they felt empowered to imagine social technologies centering joy, creativity, and authentic human connection--qualities they had once believed beyond digital spaces' reach. P10 pointed out that overcoming the design fixation around social media is crucial: \blockquote{The way that we can make social media better is if we expand the realm of what the definition of social media is. Instead of it being where someone just shares stuff, it can also be places where people connect, and people do this and that.}{10}
Several participants noted that the FI workshop played a pivotal role in helping them gain a broader vision of social technologies and envision new possibilities for social media. P15 reflected on how their perspective evolved throughout the process:
\blockquote{When we started, I literally was giving all these suggestions, which was completely the same thing which existed\ldots{} I was repeating their concepts. So I think throughout this entire interview, I evolved my idea\ldots{} \textbf{actually thinking about what is new}\ldots{} So I think this has been a super hit. I like the entire conversation.}{15}

They appreciated that SMH was not merely a \inlinequote{clone of the same thing\ldots{} without actually adding anything else} but rather a concept that was \inlinequote{completely different\ldots{} not even the same.} Similarly, P14 shared how the workshop shifted their initial skepticism: \blockquote{I don't know what else you could do, because it's online\ldots{} so I really don't know what else there could be to make it more meaningful.}{14} After engaging with the workshop, they felt empowered to imagine broader possibilities: \blockquote{I'm not really a tech person. So I don't really think about [how to make social media better] that much. But I feel like this fake scenario\ldots{} [helped] remind myself, `Girl, you have full control, this is no restrictions.' After I got into that \textbf{I\ldots{} have a way bigger vision of what social media can become.}}{14}

%% file: sections/5_discussion.tex
\section{Discussion}
This paper starts from the premise that the central challenge in reimagining social media is not a lack of ideas, but a lack of conditions that allow different ideas to emerge. Asking young people how to redesign ``social media'' is itself a loaded prompt: the term carries accumulated controversy, cynicism, and tightly constrained expectations about what platforms can and should be. As our findings show, this framing activates design fixation rather than imagination. FI offers an alternative starting point by suspending the assumptions that quietly govern what feels reasonable, feasible, or even worth proposing. FI does not magically produce ideas that have never existed. Instead, it makes different configurations of priorities and interaction logic available to think with, and legible in participants' own terms.

\subsection{Defamiliarization, Not Novelty: How FI Expands the Frontier of Social Media Design}
A useful way to interpret participants' concepts is through defamiliarization: making strange what has been naturalized so we can notice the assumptions embedded in today's dominant interaction grammar. Mainstream social media has converged on a paradigm organized around mobile-first interfaces, algorithmically curated feeds, frictionless consumption, and quantifiable performance. Under this convergence, many alternative logics are no longer legible as futures, even if their ingredients have existed before. Customizable profiles, spatial exploration, interest-based rooms, ambient co-presence, and playful identity work are not novel components. What is striking is how systematically they have been treated as side paths rather than viable organizing principles, crowded out by the priorities that dominate at scale.

``Newness'' here should not mean inventing unprecedented features. Many elements participants reach for are historically familiar: MySpace made profile customization central; Second Life organized sociality through spatial exploration; chat rooms and MUDs normalized interest-based rooms and ambient co-presence. The problem is not that these ingredients never existed, but that they have been systematically crowded out by a particular configuration that won: the mobile-optimized, algorithmically-curated, vertically-scrolling feed oriented toward frictionless consumption and legible performance. The analytic question is less ``has anyone ever built X?'' than ``why did X lose, and what does that loss reveal about what has become hard to imagine as a viable organizing principle?'' When prior platforms are invoked today, they are often treated as feature catalogs or nostalgic references, rather than as prompts to reconsider the interaction logics and priorities that once made those logics coherent.

Seen this way, FI is not a machine for generating ideas no one has ever had; it is a method for making alternative configurations thinkable again. Even if every individual component has appeared before, specific arrangements matter enormously: the smartphone, touchscreen, and app store all existed before the iPhone, but a different bundling of affordances and defaults made a new paradigm visible. Participants' proposals are revealing because they reorder emphasis: what is made reversible versus permanent, what requires consent, what counts as presence, and what the system treats as the default social unit. This reframing also helps explain why many youth examples draw from spaces not labeled ``social media'' at all: Discord servers, Minecraft worlds, and game-like environments where connection is organized around shared activity, rooms, and mutual orientation rather than broadcast audiences. FI loosens the category boundary itself, expanding what futures feel available to propose, including trajectories previously abandoned for reasons of incentive structures, scale narratives, or technical constraints rather than because their social ideas were exhausted.

FI matters because it interrupts the subtle constraints that ordinarily shape what participants feel permitted to propose. The fictional frame suspends economic, technical, and reputational pressures, but more importantly it removes the category label ``social media,'' which often triggers cynicism and incremental repair. In its place, FI invites reasoning from felt social experience: what presence should feel like, what kinds of disclosure should be low-stakes, what boundaries should be socially intelligible, and what kinds of connection should be effortless versus effortful. That shift makes it easier to surface configuration-level commitments: not whether a platform has a ``profile'' or ``messaging,'' but what counts as visibility, what is reversible, what requires consent, and what the system treats as the default social unit (an audience, a network, a room, a neighborhood, a party).

\input{inserts/diagram}
\subsection{Minecraft, Not Instagram: Interaction Logics Beyond the Feed}
Public discourse increasingly portrays social media as fundamentally detrimental to mental health, friendship, and authentic social interaction. This narrative often equates social media with specific platforms like Instagram, TikTok, or X and the ways they happen to be designed today~\cite{OtherOtherTikTokHarmst, noauthor_2023-pn}. Our study participants expressed similar perspectives, frequently describing social media as centered around superficial interactions that fail to foster meaningful connections. Yet this characterization did not match where they described actually feeling close to friends.

As illustrated in Figure \ref{fig:diagram}(a), there is a disconnect between platforms that youth commonly identify as ``social media'' (such as Instagram and TikTok) and the platforms where they actually experience meaningful connections (such as Discord or Minecraft). The latter are not typically framed as ``social media'' in everyday discourse, even though they often serve social functions more directly. This boundary is not merely semantic: it shapes what becomes imaginable. If ``social media'' is assumed to mean a 2D feed, a profile, and an attention economy, then even well-intentioned redesign prompts tend to elicit cynicism or incrementalism. When participants reasoned from spaces they already associate with social closeness (games, chat servers, shared virtual worlds), the future of social connection expanded beyond the feed metaphor.

This misalignment between expectations and experiences is problematic because negative perceptions of social media use can diminish psychological well-being~\cite{lee2024social, kim2024privacysocialnormsystematically}. Youth experience disappointment in seeking deep social connections on platforms where such interactions are scarce. As shown in Figure \ref{fig:diagram}(b), social media serves diverse uses and gratifications, from information seeking and entertainment to status enhancement and network maintenance. Rather than maintaining an entirely negative view confined to certain platforms, it is crucial to understand each platform's distinct values and limitations and to name the kinds of connection each one is organized to support.

Critiquing Instagram for not facilitating authentic, deep friendships can miss the point: while such connections might occasionally form there, it is not the platform's central form of engagement. However, platforms are not static; their designs evolve, and user engagement patterns shift over time. Instagram, primarily used for personal branding, sometimes facilitates closer communication through private `finsta' accounts. This fluidity underscores the importance of teaching~\cite{WilsonWisniewski-2012-FightingMyRegulation-m, Wisniewski-2018-PrivacyParadox-l, TheLearningNetwork-2025-WhatTeensMedia-z} youth to be more discerning about the connections they seek and what a given platform's interaction model realistically supports, rather than reinforcing anxiety-inducing narratives that treat ``social media'' as a single, uniformly harmful category.

Together, these accounts clarify the contrast participants repeatedly drew between platforms they associate with ``social media'' and the spaces where they actually feel socially connected. When describing meaningful connection, they emphasized presence, shared activity, and low-pressure co-presence, drawing on experiences in Minecraft- and Discord-like environments rather than Instagram-style broadcast and feed consumption.

\subsection{From Subjects to Authors: FI as a Translator for Youth Voice}
Youth are typically positioned in social media discourse as subjects to be protected or problems to be solved: screen time to be managed, mental health to be safeguarded, misinformation vulnerability to be addressed. This framing, however well-intentioned, reproduces a power imbalance in which young people live with the consequences of platform design earlier and more intensely than any other demographic, yet rarely get to author the underlying rules. FI offers a corrective. By treating participants as theorists and inventors rather than survey respondents or focus group informants, FI shifts the locus of imagination. The result is not merely a collection of wishes, but a disciplined translation pipeline that surfaces interaction primitives (what actions exist, what counts as visibility, what is reversible, what requires consent), normative frames (what should be easy versus hard, what should be private by default, what should be communal), and actionable design directions grounded in participants' own social experience.

This repositioning matters because asking young people to imagine within realism often reproduces the very constraints that limit design thinking. Realism already encodes corporate incentives, technical path dependencies, and fear-based narratives inherited from parents, teachers, and policymakers. When participants are prompted to ``fix'' social media, they inherit a frame saturated with cynicism and incrementalism. FI sidesteps this inheritance by suspending the category label ``social media'' altogether. Crucially, the fictional frame does more than remove constraints; it makes certain desires and boundaries socially permissible to articulate. Wishes that might feel naive, awkward, or excessively earnest under a realistic prompt—wanting connection to feel like being in the same room, wanting visibility to be reversible, wanting strangers to remain strangers unless actively invited—become speakable when the task is to describe a magical world rather than repair an existing platform. In place of defensiveness, participants reason from felt social experience: what presence should feel like, what kinds of disclosure should be low-stakes, what boundaries should be socially intelligible.

A practical consequence is that FI generates \textit{common language} between youth and adults. Much of the current tension in public discourse around social media stems from a mismatch in vocabulary: adults speak in terms of harm, addiction, and regulation, while youth often lack the conceptual tools to articulate what they want connection to feel like. FI bridges this gap by making youth priorities legible in terms that designers, researchers, and policymakers can act upon. When participants describe wanting ``rooms'' instead of ``feeds,'' or ``neighborhoods'' instead of ``audiences,'' they are not simply expressing preference for a feature; they are naming a different interaction grammar. That grammar, once articulated, becomes available for translation into design specifications, policy frameworks, and educational curricula. The contribution is not a single new feature, but evidence that a different organizing logic is possible and legible to users when the design frame allows it.

Finally, FI cultivates something often missing from conversations about social media: hope. The shift we observed from ``I do not know what else there could be'' to expansive, imaginative visions is not incidental. It reflects what happens when participants are given back authorship over the future, even temporarily. Past platforms were designed by adults with particular business models and assumptions baked in; youth imagining social worlds from scratch surfaces different priorities. The deeper problem is not that no one has tried to improve social media, but that most attempts operate within the same paradigm: adjusting privacy settings, adding screen time limits, tweaking algorithmic recommendations. These are repairs to a house whose floor plan is assumed fixed. FI invites participants to redraw the floor plan itself, supporting not only imagination but critical capacity: when youth can name what they want connection to feel like, and can describe the rules that would make that possible, they are better positioned to evaluate platforms critically without being trapped in cynicism. FI supports youth voice not only by eliciting ideas, but by cultivating the sense that different interaction futures are thinkable and therefore arguable.

\subsection{Limitations and Future Work}
This study is exploratory, and the design orientations we synthesize should be understood as generative rather than generalizable. Our participant demographics skewed toward women, and future work should examine how robust these orientations are across broader samples and cultural contexts, including youth with different relationships to gaming, spatial interfaces, and platform norms. Also, FI intentionally suspends feasibility and market constraints, which is a strength for defamiliarization but a limitation for direct translation into deployable systems. Some of the most compelling interaction logics we surfaced (spatially legible boundaries, ambient co-presence, rich memory-sharing) raise practical questions about accessibility, moderation, harassment, surveillance, and inequities in device access. Future work should prototype selectively, using low-fidelity and mid-fidelity implementations to evaluate whether the same felt qualities can be achieved under real constraints, and to identify new risks that emerge when imagined systems become operational.

%% file: inserts/diagram.tex

\begin{figure}[t]
    \centering
    \includegraphics[width=0.5\linewidth]{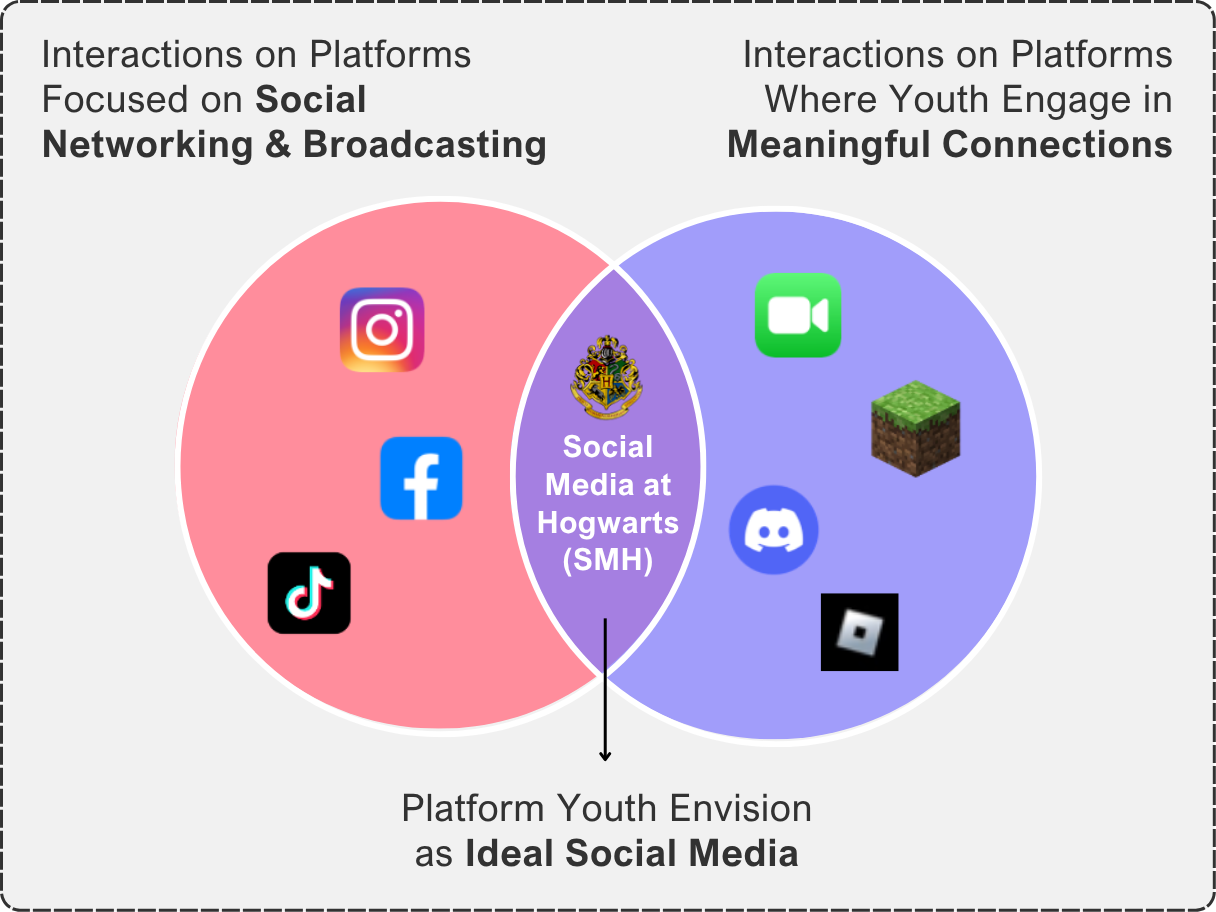}
    \caption{Youth's Understanding of Social Media Interactions}
    \Description{A Venn diagram comparing two categories of online platforms. The left circle represents platforms focused on social networking and broadcasting, with icons for Instagram, Facebook, and TikTok. The right circle represents platforms where youth engage in meaningful connections, with icons for Zoom, Minecraft, Discord, and Roblox. In the overlapping center, labeled ``Social Media at Hogwarts (SMH)'', the diagram highlights youth's vision of an ideal platform that combines the social visibility of networking platforms with the depth of meaningful connection found in gaming and community spaces.}
    \label{fig:diagram}
\end{figure}

%% file: sections/6_conclusion.tex
\section{Conclusion}

This paper demonstrates that when youth are invited to imagine social connection outside the conceptual frame of ``social media,'' they articulate desires and interaction logics that diverge sharply from dominant platform paradigms. Through FI workshops with 23 participants aged 15--24, we surfaced six themes---intuitive social navigation, shared experiences, communal ambience, nuanced self-presentation, intentional engagement, and playfulness---that collectively describe what meaningful remote connection might feel like if it were organized around presence, place, and relationships rather than feeds, metrics, and passive consumption.

The findings carry implications for both research and practice. For researchers, this study offers evidence that design fixation in social media is partly a methodological problem: the term ``social media'' itself activates assumptions that constrain imagination. FI provides one way to interrupt those assumptions, enabling participants to reason from felt social experience rather than inherited platform templates. For designers and policymakers, the themes we identify suggest that the relational shortcomings of mainstream platforms are not inevitable features of mediated connection but outcomes of specific design choices---choices that could be made differently.


%% file: sections/7_appendix.tex
\clearpage
\appendix
\label{appendix}

\section{Fictional Inquiry Materials}
\label{sec:fi-materials}

\subsection{Miro Board for Fictional Inquiry Warm-Up}
\label{sec:miro-appendix}
\input{inserts/miro}
Figure~\ref{fig:miro} shows the Miro board template used during Fictional Inquiry sessions. Participants sorted their friends and family into Potter-inspired categories (houses, teams, old friends) to encourage thinking about relationships within the fictional context and enhance immersion in the design space. This activity preceded the main design exploration and helped participants transition from discussing their current social media experiences to imagining alternatives within the Hogwarts setting.

\section{FI Workshop Protocol}
\label{protocol}
\input{inserts/protocol}

%% file: inserts/miro.tex

\begin{figure}[t]
    \centering
    \includegraphics[width=0.9\linewidth]{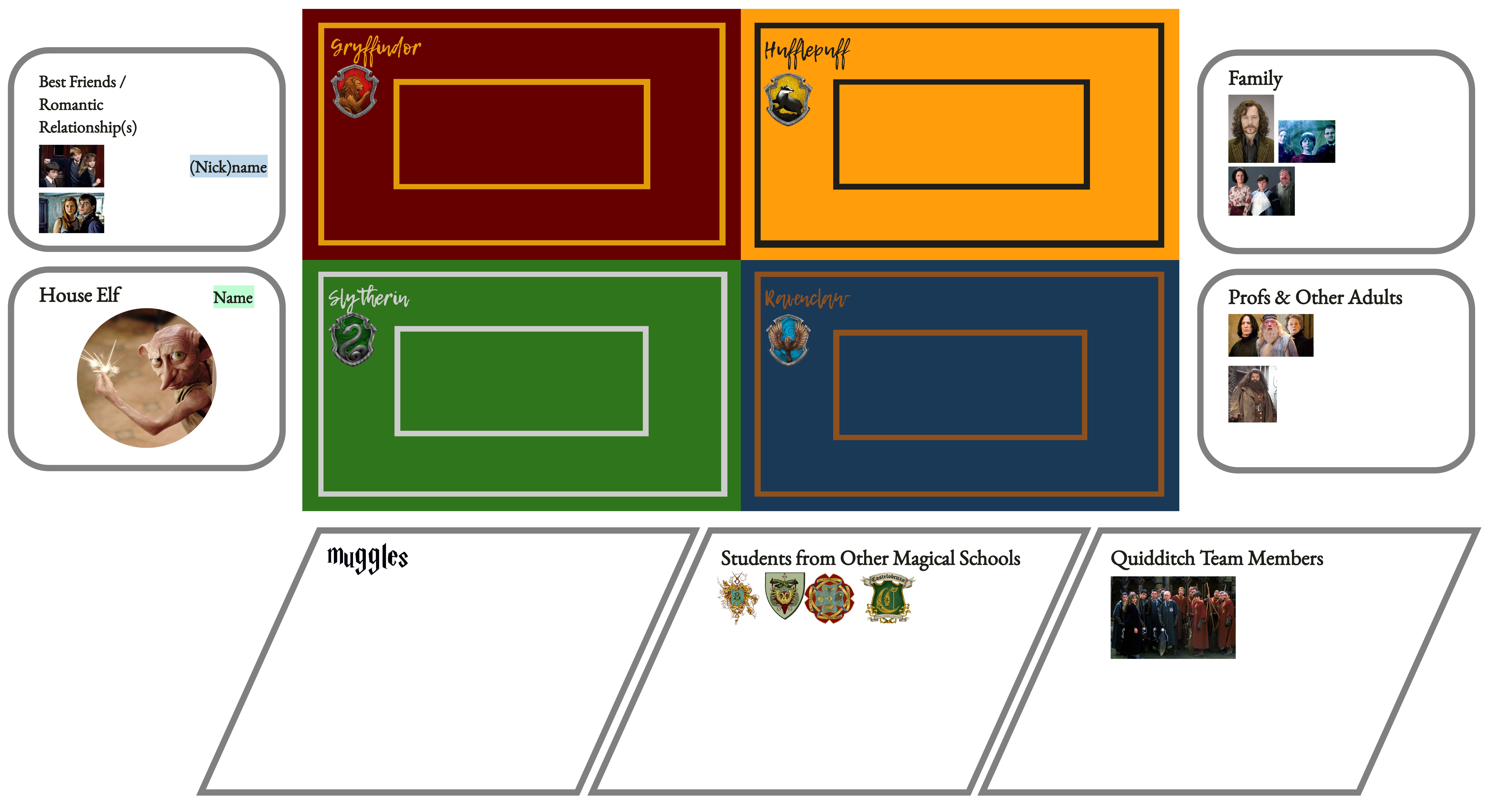}
    \caption{Miro board mapping real-life relationships to Hogwarts-themed groups, designed to immerse participants in the Hogwarts-inspired FI narrative.}
    \Description {This is a visual chart designed to organize relationships and connections in the Harry Potter universe. There are three columns of labeled sections. In the middle, there are four sections, each representing one of the Hogwarts houses: Gryffindor (red), Hufflepuff (yellow), Slytherin (green), and Ravenclaw (blue). Each house box has its crest. To the left, there are two smaller boxes. The top-left box is labeled ``Best Friends / Romantic Relationship(s),'' featuring small images of Harry Potter, Hermione Granger, and Ron Weasley, with a space for a name or nickname. Below this, there is a smaller box labeled ``House Elf,'' featuring a picture of Dobby holding a glowing object and an area for a name. 
    To the right, there are two more boxes. The top-right box is labeld ``Family,'' containing images of Sirius Black, James and Lily Potter, and the Dursley family. Below this, another box is labeld ``Profs \& Other Adults,'' displaying images of Severus Snape, Dumbledore, Professor McGonagall, and Hagrid.
    At the bottom, there are three large sections. The leftmost section is labeled ``Muggles'' with a blank space.
    The middle section, labeld ``Students from Other magical Schools,'' displays crests from Beauxbatons, Durmstrang, Ilvermorny, and Castelobruxo.
    The rightmost section, labeled ``Quidditch Team Members,'' features an image of a Quidditch team in red robes.
    Each section contains an empty space for names or details to be added.}
    \label{fig:miro}
\end{figure}


%% file: inserts/protocol.tex

\subsection*{Introduction}
Welcome, and thank you for participating in our design workshop. Before we begin, I want to ensure that you've read the consent form I sent earlier. Have you had a chance to review it?

\textbf{[Response]}

\noindent Great! Just to reiterate, this session will be recorded for research purposes. Do I have your consent to proceed with the recording?

\textbf{[Response]}

\noindent Thank you. If at any point you feel uncomfortable or wish to stop the interview, please let me know.

\subsection*{General}
Let's start with some general questions about your experiences with 3D gaming and Harry Potter.

\begin{itemize}
    \item Which 3D games do you play and how long have you been playing them?
    \item (Which is your favorite?) Can you describe the 3D elements or the spatial elements in that game a little bit, please?
    \item Have you ever played in VR mode if that is available?
    \item Any AR games you have played? Like Pokémon Go?
    \item What Harry Potter books or movies have you watched?
    \item Which is your favorite?
    \item Which Hogwarts house would you put yourself in?
\end{itemize}

\subsection*{Narrative/Plot}
As mentioned in the consent form, our goal today is to explore innovative ways to improve social media. We particularly invited 3D game players with a love for Harry Potter because we believe this will help us think creatively and ``outside the box'' about what an ideal, magical social media could look like, beyond the confines of a small 2D smartphone screen.

Don't worry about being too creative---we're here to guide you through the design process, not to evaluate you. We want to explore different design directions with your help, and we'll be actively involved while also giving you the space to share your ideas.

To give our brainstorming some structure, we'll start with a specific scenario to solve during this session. I'm going to ask you some guiding questions to help you address the scenario:

\begin{quote}
\textit{Imagine you are a student at Hogwarts. You have friends who are Muggle-born, friends from magical families, and even friends at other wizarding schools around the world. You're also getting to know me, as I'm a new student you've just met at Hogwarts. Keeping in touch with everyone is tough because traditional Muggle social media like Instagram doesn't really portray your wizard self very well. You don't have a smartphone in the first place because Hogwarts people don't need electronic devices, so you can't really charge your smartphone.}\\
\textit{One day, a brilliant Muggle-born student had an idea. They decided to use magic to create a magical social media where everyone---Muggles, Hogwarts students, professors, and friends from other magical schools---could all communicate and share their lives in the most ideal way. Imagine you're using this new magical platform---or maybe it's not even a platform. Whatever it is, what do you think this would look like?}
\end{quote}

\subsection*{Character Map}
Before we dive into the designing part, let's first map out your relationships in your new life as a Hogwarts student. Use the Miro board link (Figure \ref{fig:miro}) to write down the names or nicknames of people that you would like to put in each section. For example, you might want to put ``younger sister'' or ``Jane'' under Family, some of your coworkers under Quidditch Team members, yourself under the house that you think you belong to, your older middle school friends as Muggles, etc.

Please note the two boundaries in the four rectangles for each house. You can place your closer friends in the inner box and not-so-close ones in the outer box. And you can put the same name in multiple boxes (or parallelograms) as you wish.

\begin{itemize}
    \item How do you currently communicate with or stay connected with each of these people via Muggle social media?
    \item What are your biggest pet peeves/issues with Muggle social media that you'd like to solve with magical powers?
    \item What moments do you feel bad/guilty about using social media, if at all?
    \item When does time spent on social media feel meaningless/meaningful/fulfilling, if at all?
    \item Pet peeves around privacy settings?
    \item When do you feel most connected to your friends when interacting with them on social media?
    \item What would a better social media be if you were to use adjectives to describe it? What does ``better'' mean to you?
\end{itemize}

\subsection*{Design}
Now, let's think about what this magical world social media might look like. Remember, there are no right or wrong answers here. We're looking for your imagination and creativity, so feel free to think outside the box and have fun with it!

\begin{itemize}
    \item What key capabilities would this thing have for communicating with [GROUP]*?
    \item If you could design this thing with any magical powers, what form would it take? (Is it a magical mushroom? A secret room? A creature?)
    \item What would you want to share about yourself on this ideal magical ``thing'' with [GROUP]?
    \item If you had [GROUP'] join this ``thing,'' how might its form adapt to include this new group?
\end{itemize}

*: Iterate for 1) Romantic Relationships and Closest Friends, 2) Closer Hogwarts Friends, 3) Muggle Friends, 4) Friends at Other Magical Schools, 5) Quidditch Team Members, 6) House Elf

\subsection*{Connecting Back}
\begin{itemize}
    \item What do you like about this magical ``thing'' compared to Muggle social media?
    \item What do you dislike about it?
    \item Imagine using the magical ``thing'' we discussed earlier. How do you think it could resolve some of the issues people might face with Muggle social media?
    \item How might this new magical ``thing'' better support your needs for self-presentation and identity management compared to traditional social media?
    \item What challenges might arise in using such a ``thing'' at Hogwarts?
\end{itemize}

Thank you so much for your time and insights. Your contributions are incredibly valuable to our research. Do you have any final thoughts or questions before we conclude?
